\documentclass[11pt]{article}
\usepackage[margin=2cm]{geometry}
\usepackage{cite}
\usepackage{geometry}                
\usepackage{graphicx}
\usepackage{amssymb}
\usepackage{amsmath,amssymb}
\usepackage{epsfig}
\usepackage{graphicx}
\usepackage[normalem]{ulem}

\usepackage[usenames,dvipsnames]{color}

\def\Tr{\mathrm{Tr}} 

\title{
\vspace*{-2cm}
\begin{flushright}
\normalsize{EFI-12-33\\
ANL-HEP-PR-12-101}
~\\
\end{flushright}
\vspace*{1.5cm}
$SU(2)\otimes SU(2)$ Gauge Extensions of the MSSM Revisited\\
\author{\textbf{Ran Huo$^a$, Gabriel Lee$^a$, Arun M. Thalapillil$^{a,d}$ and Carlos E.M. Wagner$^{a,b,c}$} \\
~\\
\normalsize\emph{$^a$Enrico Fermi Institute, University of Chicago, Chicago, IL 60637}\\
\normalsize\emph{$^b$HEP Division, Argonne National Laboratory, 9700 Cass Ave., Argonne, IL 60439}\\
\normalsize\emph{$^c$Kavli Institute for Cosmological Physics, University of Chicago, Chicago, IL 60637}\\
\normalsize\emph{$^d$Department of Physics, Rutgers University, Piscataway, NJ 08854} \\
}}

\begin{document}
\maketitle
\vspace*{0.5cm}
\begin{abstract}
We study an extension of the Minimal Supersymmetric Standard Model with a gauge group $SU(2)_1\otimes SU(2)_2$ breaking to $SU(2)_L$. The extra wino has an enhanced gauge coupling to the SM-like Higgs boson and, if light, has a relevant impact on the weak scale phenomenology.
The low energy Higgs quartic coupling is modified both by extra $D$-term corrections and by a modification of its renormalization group evolution from high energies. At low values of $\tan\beta$,  the latter effect may be dominant.  This leads to interesting regions of parameter space in which the model can accommodate a 125 GeV Higgs with relatively light third generation squarks and an increased $h \rightarrow \gamma \gamma$ decay branching ratio, while still satisfying the constraints from electroweak precision data and Higgs vacuum stability.
\end{abstract}
\thispagestyle{empty}


\section{Introduction}
The LHC experiments have recently reported the discovery of a new particle, with properties similar to the ones of the  Standard Model (SM) Higgs boson, and with a mass of roughly 125 GeV \cite{ATLAS:2012ae,ATLAS:2012gk,Chatrchyan:2012gu,Chatrchyan:2012tx}. This Higgs-like particle has been detected in its decay into a pair of photons or massive gauge bosons, $ZZ^*$ and $WW^*$, with a rate which is roughly consistent with the one expected in the SM.   The fact that this Higgs-like particle is produced via gluon fusion provides indirect evidence of its coupling to top-quarks. The coupling to other fermions is uncertain, although the recent LHC and Tevatron data have provided some evidence of its couplings to bottom-quarks and $\tau$ leptons, at a rate consistent with the SM one~\cite{Aaltonen:2012qt,LHCbottomtau}.

A SM-like Higgs boson, with a mass of about 125~GeV,  provides constraints on the parameters of  the SM and its possible extensions.  In the SM, such a Higgs boson mass fixes the low energy value of the quartic coupling of the Higgs potential, which is otherwise a free parameter. In the Minimal Supersymmetric Extension of the SM (MSSM), instead,  the effective quartic coupling of the SM-like tree-level Higgs potential in the limit of large non-standard Higgs boson masses is related to the gauge couplings by the supersymmetric D-terms. Consequently, at tree level, the mass of the SM-like Higgs boson is bounded to be below the $Z$ boson mass, and a large loop correction is needed to achieve a Higgs mass of 125 GeV. This large loop correction may be provided by quantum corrections induced by the third generation quarks and their supersymmetric partners. It has long been established  that stop masses of order 1~TeV can lead to a SM-like Higgs as heavy as 130~GeV, provided $\tan\beta$, the ratio of the two Higgs boson vacuum expectation values, is large and the left-right stop mixing parameter $A_t$ is of the order of 2.4~times the average stop mass~\cite{Higgstwoloops}. Values of the SM-like Higgs mass of about 125~GeV  require values of the parameters close to the ones leading to this maximal value, namely moderate to large $\tan\beta$ to maximize the tree-level mass, at least one heavy stop enabling sizable top quark-loop contributions to the renormalization group evolution (RGE) of the quartic coupling, and a large stop mixing parameter $\tilde{A}_t$ of the order of, or larger than, the heaviest stop mass (see for example   \cite{Hall:2011aa, Arbey:2011ab,  Draper:2011aa,Carena:2011aa} and related references).
\par
Interestingly, the current best fit for the rate of the diphoton production channel, proceeding from Higgs production and decay, suggests an enhancement  over the SM expectation~\cite{ATLAS:2012ae,ATLAS:2012gk,Chatrchyan:2012gu,Chatrchyan:2012tx},
\begin{equation}
\frac{\left[\sigma(pp\rightarrow h)\times BR(h\rightarrow\gamma\gamma)\right]_{\text{best-fit}}}{\sigma_{\text{SM}}(pp\rightarrow h)\times BR_{\text{SM}}(h\rightarrow\gamma\gamma)}  =
1.9 \pm 0.5, \;\;  (1.56 \pm 0.43)
\end{equation}
at the ATLAS (CMS) experiment; however, the errors are still too large to claim any significant deviation with respect to the SM.  On the other hand, the preliminary indications are that $h\rightarrow Z Z^*$ and $h\rightarrow W W^*$ are in better alignment with the SM expectations.

Although it is premature to draw any firm conclusion from the present Higgs data, it is interesting to consider what would be the consequences of an enhanced diphoton decay rate. Since this rate is induced at the loop level, new weak scale charged particles can induce a modification of this rate, and therefore  a lot of effort has been directed towards understanding and incorporating this enhanced diphoton branching ratio (BR) in weak scale extensions of the SM~\cite{Djouadi:1998az, hgagaRef1,hgagaRef2, hgagaRef3, dnq, Carena:2012mw, Low:2009di, Low:2009nj, Carena:2012xa, Carena:2012gp, Joglekar:2012vc, Blum:2012ii}.
\par
The enhancement of the Higgs diphoton decay rate may be induced by particles carrying not only charge but also color.  Color charged particles will lead not only to a modification of the Higgs diphoton decay rate but also to the main gluon fusion production mode.  In the MSSM, for instance, a light stop $\tilde{t}$ or other colored particles may enhance the Higgs diphoton decay BR, but they will also simultaneously lower the Higgs gluon-fusion production cross section. 
This is hard to accommodate in view of the current data\cite{ATLAS:2012ae,ATLAS:2012gk,Chatrchyan:2012gu,Chatrchyan:2012tx} and other theoretical constraints~\cite{Reece:2012gi}.  We shall therefore focus on the scenario that the particles carry no color and there is an enhancement of the Higgs diphoton decay BR compared with the SM, without any relevant modification of the main gluon fusion and weak boson fusion Higgs production rates.
Moreover, we shall consider the case in which the main decay modes into massive gauge bosons and bottom quarks remain close to the SM.  In supersymmetric extensions, this is achieved by a sizable value of the non-standard Higgs boson mass $M_A$, of the order of, or larger than, 1~TeV. 

In this paper, we explore a gauge extension of the MSSM that may simultaneously account for the 125 GeV SM-like Higgs boson and also yield an enhancement of the Higgs diphoton decay BR.\footnote{An extension of the MSSM by an $SU(2)_L$ triplet with $\mathcal{O}(1)$ coupling to the Higgs, in a context without gauge extensions, was explored in \cite{dnq}.} The model is based on \cite{Batra:2003nj,Batra:2004vc} (henceforth BDKT): the gauge sector is extended to $SU(2)_1\otimes SU(2)_2$ at a high scale ($SU(2)_2$ is asymptotically free and its coupling becomes large at EW scales) which is spontaneously broken to the SM $SU(2)_L$ by the vacuum expectation value (vev) of a bidoublet $\Sigma$. In the original study, at low energy, the propagator of the residual scalar component of the bidoublet gives an extra effective contribution to the supersymmetric D-term, and consequently an extra contribution to the tree-level Higgs mass,  as well as to the mass splitting of the isospin components of those scalar fields transforming in a non-trivial way under SU(2)~\cite{Medina:2009ey}.
\par
We investigate a specific aspect of this model that was previously overlooked. In certain regions of the parameter space, the lightest wino has a predominant component from the second, strongly interacting $SU(2)_2$.  These light charginos couple strongly to the SM-like Higgs and may lead to a relevant modification of the Higgs diphoton decay BR with respect to the SM one. In the MSSM, the enhancement of the diphoton rate mediated by charginos is small due to the weak couplings of the Higgs to the charginos.  In this case, however, the chargino effects are enhanced by a factor $(g_2/g)^2$ with respect to the ones in the MSSM, where $g_2$ is the $SU(2)_2$ coupling and $g$ is the SM weak coupling.  The coupling $g_2$ cannot be arbitrarily large, however, since it would induce a large correction to the Higgs mass beyond the D-term modifications. This is a new, unexplored mechanism for the modification of the Higgs mass, which goes beyond the frameworks of non-decoupling D-terms  discussed in \cite{Blum:2012ii}, and therefore will remain interesting even in the case that the diphoton enhancement were not present after analyzing the full set of experimental data.  We will explore viable regions of parameter space where this model is consistent with current experimental and theoretical constraints. 
\par
The paper is organized as the following. We review the model in detail in section 2. In section 3, we discuss the phenomenology of the model. Numerical results are presented in section 4. We present our conclusions in section 5.


\section{$SU(2)_1 \otimes SU(2)_2$ Extension of the MSSM}

At high scales, the extended weak gauge sector is $SU(2)_1\otimes SU(2)_2$. The $SU(2)_1$ vector supermultiplet couples only to the first two generations of chiral supermultiplets in the MSSM, while the $SU(2)_2$ vector supermultiplet couples only to the third generation chiral supermultiplet and the two Higgs supermultiplets. Such an arrangement will lead to vanishing tree-level couplings of the Higgs to the first two generations due to gauge invariance, so a natural hierarchy of masses between the heavy and light generations appears in this model. Due to the asymptotically free nature of $SU(2)_2$, at the weak scale, the related gauge couplings satisfy $g_2 > g_1$. A bidoublet chiral supermultiplet $\Sigma$ with the charge $(\mathbf{2,\bar{2}})$ under $SU(2)_1\otimes SU(2)_2$ links the two gauge group sectors. We can represent the bidoublet as a matrix
\begin{equation}
\Sigma=\Sigma^0\mathbf{1}+\Sigma^a\sigma^a,
\end{equation}
where $\sigma^a$ are the Pauli matrices. There is also a gauge singlet  field $S$, which provides supersymmetric masses to the bidoublet fermion and scalar fields. With these ingredients, the effective superpotential of the model is given by
\begin{equation}\label{superpotential}
\mathcal{W}=\mathcal{W}_{{\scriptsize\mbox{MSSM}}}-\lambda S\Big(\frac{1}{4} \Sigma\Sigma -\omega^2 \Big) + \frac{\kappa}3 S^3.
\end{equation}
Note that $\Sigma\Sigma$ is contracted with two epsilon tensors, i.e. $\Sigma\Sigma = 2 \det (\Sigma) = 2((\Sigma^{0})^2-\vec{\Sigma}^2)$. We assume $\kappa \ll \lambda$, and with the addition of a soft mass term for $S$, $S$ acquires a large vev $\langle S\rangle \sim \mathcal{O}$(TeV$/\kappa$). The potential relevant to the $\Sigma$ field is, from above,
\begin{equation}\label{potential}
V_\Sigma = \frac{1}{16} |\lambda|^2|\Sigma\Sigma|^2 - \frac{1}{4}|\lambda|^2\omega^2\Big(\Sigma\Sigma+\mbox{h.c.}\Big) + {\frac{1}{2}}\Big( |\lambda|^2 \langle S^\ast S \rangle +m_\Sigma'^2 \Big) \Tr \big(\Sigma^\dag\Sigma\big) + \textrm{D-terms} \, ,
\end{equation}
where $m_\Sigma^{\prime}$ is the soft mass term corresponding to the scalar component of the $\Sigma$ supermultiplet, and, for simplicity, we have ignored subdominant terms of ${\cal{O}}(\kappa)$. Observe that the fermionic component $\tilde{\Sigma}$ obtains the mass $M_{\tilde{\Sigma}}=\lambda\langle S\rangle$ from the Yukawa terms, where we have chosen a different normalization of the coupling $\lambda$  from the one in \cite{Batra:2003nj,Batra:2004vc}. The total mass of the scalars is defined as $m_\Sigma^2=M_{\tilde{\Sigma}}^\dag M_{\tilde{\Sigma}}+m_\Sigma'^2$.

\subsection{$SU(2)_1\otimes SU(2)_2\rightarrow SU(2)_L$ Spontaneous Symmetry Breaking}
The potential in Eq. (\ref{potential}) will induce spontaneous symmetry breaking and result in
\begin{equation}
SU(2)_1\otimes SU(2)_2\rightarrow SU(2)_L.
\end{equation}
We choose a diagonal breaking, with $\Sigma$ acquiring a vev of the form $\langle\Sigma\rangle=u\mathbf{1}$. Minimizing the potential, we find
\begin{equation}
u^2= 2\Big(\omega^2-\frac{m_\Sigma^2}{|\lambda|^2}\Big).
\end{equation}
Analogous to electroweak symmetry breaking (EWSB), the gauge bosons $W_1$ and $W_2$ of $SU(2)_1 \otimes SU(2)_2$ mix to form $W'$ bosons, which have masses proportional to $u$, and the SM $W$ bosons, which remain massless, leading to the unbroken $SU(2)_L$. At lower energies, these $W$ bosons and the $U(1)_Y$ gauge boson $B$ combine after EWSB to form the photon and the massive $W^\pm$ and $Z^0$ bosons.
\par
In analogy with chiral perturbation theory, the covariant derivative of the $(\mathbf{2,\bar{2}})$ bidoublet is
\begin{equation}
D_\mu\Sigma=\partial_\mu\Sigma+ig_1W_{1\mu}^at^a\Sigma-ig_2\Sigma W_{2\mu}^at^a,
\end{equation}
where $t^a=\frac{1}{2}\sigma^a$ are the $SU(2)$ generators. The kinetic term for the scalar component of the bidoublet is
\begin{equation}
\frac{1}{2} \mathrm{Tr} ((D^\mu \Sigma)^\dag D_\mu \Sigma).
\end{equation}
Inserting the vev $u$ for the scalar, we obtain the following mass term for the $W'$ gauge bosons:
\begin{equation}
\frac{u^2}{4} (g_1W_1^a-g_2W_2^a)^2.
\end{equation}
The mixing matrices between the gauge eigenstates ($W_1$, $W_2$) and the mass eigenstates ($W$, $W'$) are
\begin{equation}\label{Wmixing}
\left(\begin{array}{c}
W' \\
W
\end{array}\right)=\left(\begin{array}{cc}
\frac{g_1}{\sqrt{g_1^2+g_2^2}} & -\frac{g_2}{\sqrt{g_1^2+g_2^2}} \\
\frac{g_2}{\sqrt{g_1^2+g_2^2}} & \frac{g_1}{\sqrt{g_1^2+g_2^2}}
\end{array}\right)\left(\begin{array}{c}
W_1 \\
W_2
\end{array}\right).
\end{equation}
From above, we note that the $W'$ bosons obtain masses $m_{W'} = \sqrt{\frac{1}{2}(g_1^2+g_2^2)}u$. Plugging these mass eigenstates into the covariant derivatives of the MSSM chiral supermultiplets, we obtain the covariant derivatives for the third generation fermions and the Higgs sectors (which as mentioned before are only charged under the $SU(2)_2$)
\begin{equation}\label{W3rd}
D_\mu=\partial_\mu+ig_2W_{2\mu}^at^a=\partial_\mu+i\frac{g_1g_2}{\sqrt{g_1^2+g_2^2}}W_\mu^at^a-i\frac{g_2^2}{\sqrt{g_1^2+g_2^2}}W_\mu^{\prime a}t^a=\partial_\mu+igW_\mu^at^a-i\frac{g_2}{g_1}gW_\mu^{\prime a}t^a ,
\end{equation}
and for the first and second generation fermions (which are only charged under the $SU(2)_1$)
\begin{equation}\label{W1st2nd}
D_\mu=\partial_\mu+ig_1W_{1\mu}^at^a=\partial_\mu+i\frac{g_1g_2}{\sqrt{g_1^2+g_2^2}}W_\mu^at^a+i\frac{g_1^2}{\sqrt{g_1^2+g_2^2}}W_\mu^{\prime a}t^a=\partial_\mu+igW_\mu^at^a+i\frac{g_1}{g_2}gW_\mu^{\prime a}t^a .
\end{equation}
Here, we have identified
\begin{equation}
g=g_{\text{SM}}\equiv\frac{g_1g_2}{\sqrt{g_1^2+g_2^2}}\, ,
\end{equation}
or equivalently $1/g^2=1/g_1^2+1/g_2^2$, which is the usual electroweak coupling constant at low energies. Note that all fermion generations have the same SM coupling (i.e. $g_{\text{SM}}$) to $W$ bosons as required by gauge invariance. 


\subsection{The Extended Chargino and Neutralino Sector}
We now turn to the chargino and neutralino sectors. Along with the bino and Higgsino, we have two sets of winos and the fermionic component of $\Sigma$. Thus, we have 4 charginos, which we denote in the gauge eigenstate basis as $(\tilde{\chi}^\pm)^T\thinspace=\thinspace(\tilde{W}_1^\pm,\thinspace\tilde{W}_2^\pm,\thinspace\tilde{H}_{u/d}^\pm,\thinspace\tilde{\Sigma}^\pm)$. Similarly, there are 6 neutralinos of interest in our model, which we denote in the gauge eigenstate basis as $(\tilde{\chi}^0)^T\thinspace=\thinspace(\tilde{B}^0,\thinspace\tilde{W}_1^0,\thinspace\tilde{W}_2^0,\thinspace\tilde{H}_d^0,\thinspace\tilde{H}_u^0,\thinspace\tilde{\Sigma}^3)$.\footnote{We can also use the basis with ($\tilde{W}_1, \tilde{W}_2$) rotated to ($\tilde{W}, \tilde{W}')$ in the same way as their bosonic counterparts, but that will introduce mixing terms between the $\tilde{W}$ and $\tilde{W}'$ in the chargino and neutralino mass matrix.}


Observe that since only the second $SU(2)_2$ couples with the Higgs sector, the $\tilde{W}_2$ Higgsino mass mixing entries have a coupling $g_2$ instead of $g$. Both the $\tilde{W}_1$ and $\tilde{W}_2$ couple to the bidoublet through the scalar-fermion-gaugino term
\begin{equation}\label{w1w2bicpl}
\mp\frac{1}{\sqrt{2}}g_i\mbox{Tr}(\Sigma^\dag\tilde{W}^a_it^a\tilde{\Sigma})+\text{ h.c.}=\mp \frac{1}{\sqrt{2}} g_iu\thinspace(\tilde{\Sigma}^-\tilde{W}^+_i+\tilde{W}^-_i\tilde{\Sigma}^++\tilde{W}^3_i\tilde{\Sigma}^3)+\text{ h.c.}
\end{equation}
Here the negative (positive) sign is for $i = 1$ (2).


We can write the mass term for the charginos in the Lagrangian as $\mathcal{L} \ni -(\tilde{\chi}^{-}_G)^T M^\pm \tilde{\chi}^{+}_G + \text{h.c.}$, where $M^\pm$ is the extended chargino mass matrix
\begin{eqnarray}\label{Chargino12}
M^\pm_{ij} &=&\left(\begin{array}{cccc}
M_{\tilde{W}_1} & 0 & 0 & \frac{1}{\sqrt{2}}g_1u \\
0 & M_{\tilde{W}_2} & \frac{1}{\sqrt{2}}g_2vs_\beta & -\frac{1}{\sqrt{2}}g_2u \\
0 & \frac{1}{\sqrt{2}}g_2vc_\beta & \mu & 0 \\
\frac{1}{\sqrt{2}}g_1u & -\frac{1}{\sqrt{2}}g_2u & 0 & M_{\tilde{\Sigma}} \\
\end{array}\right),
\end{eqnarray}
with $(\tilde{\chi}^{-}_G)^T=\thinspace(\tilde{W}^-_1,\thinspace\tilde{W}^-_2,\thinspace\tilde{H}^-_d,\thinspace\tilde{\Sigma}^-)$ and $\tilde{\chi}^{+}_G =\thinspace(\tilde{W}^+_1,\thinspace\tilde{W}^{+}_2,\thinspace\tilde{H}^+_u,\thinspace\tilde{\Sigma}^+)^T$ the charginos written in gauge eigenstates (note the subscript $G$).
The two soft masses of $\tilde{W}_1$ and $\tilde{W}_2$ are $M_{\tilde{W}_1}$ and $M_{\tilde{W}_2}$, respectively.
Similarly, we can write the mass term for the neutralinos as $\mathcal{L} \ni -\frac12 (\tilde{\chi}^{0}_G)^T M^0 \tilde{\chi}^{0}_G + \text{h.c.}$, where $M^0$ is the extended neutralino mass mixing matrix
\begin{eqnarray}\label{Neutralino12}
M^0_{ij}=&\left(\begin{array}{cccccc}
M_1 & 0 & 0 & -\frac{1}{2}g'vc_\beta & \frac{1}{2}g'vs_\beta & 0 \\
0 & M_{\tilde{W}_1} & 0 & 0 & 0 & \frac{1}{\sqrt{2}}g_1u \\
0 & 0 & M_{\tilde{W}_2} & -\frac{1}{2}g_2vc_\beta & \frac{1}{2}g_2vs_\beta & -\frac{1}{\sqrt{2}}g_2u \\
-\frac{1}{2}g'vc_\beta & 0 & -\frac{1}{2}g_2vc_\beta & 0 & -\mu & 0 \\
\frac{1}{2}g'vs_\beta & 0 & \frac{1}{2}g_2vs_\beta & -\mu & 0 & 0 \\
0 & \frac{1}{\sqrt{2}}g_1u & -\frac{1}{\sqrt{2}}g_2u & 0 & 0 & M_{\tilde{\Sigma}}
\end{array}\right) ,
\end{eqnarray}
and $\tilde{\chi}^{0}_G \thinspace=\thinspace(\tilde{B},\thinspace\tilde{W}^3_1,\thinspace\tilde{W}^{3}_2,\thinspace\tilde{H}^0_d,\thinspace\tilde{H}^0_u, \thinspace\tilde{\Sigma}^3)^T$. We have omitted the $\tilde{\Sigma}^0$, which acquires a mass $M_{\tilde{\Sigma}}$ and does not mix with the other fields, and the singlino with mass $\kappa \langle S \rangle$ that mixes with $\tilde{\Sigma}^0$ by terms of the order $\lambda u$.

As in the MSSM, we diagonalize the mass matrices to obtain the physical chargino and neutralino states. The asymmetric chargino mixing matrix is diagonalized by extended $4\times4$ unitary matrices $\mathcal{U}$ and $\mathcal{V}$. They are defined as
\begin{equation*}
(\tilde{\chi}^{-}_G)^TM_{\tilde{\chi}^\pm}\tilde{\chi}^{+}_G=\big((\tilde{\chi}^{-}_G)^T\mathcal{U}^\dag\big)\big(\mathcal{U}M_{\tilde{\chi}^\pm}\mathcal{V}^\dag\big)\big(\mathcal{V}\tilde{\chi}^{+}_G\big)
=(\tilde{\chi}^{-}_M)^T\mbox{diag}(M_{\tilde{\chi}^\pm})\tilde{\chi}^{+}_M ,
\end{equation*}
\begin{eqnarray}
(\tilde{\chi}^{-}_M)_i=\mathcal{U}_{ij} (\tilde{\chi}^{-}_G)_j , & \qquad & (\tilde{\chi}^{+}_M)_i=\mathcal{V}_{ij} (\tilde{\chi}^{+}_G)_j  .
\end{eqnarray}
The subscript $M$ denotes mass eigenstates.
\par
In the neutralino sector, the symmetric neutralino mixing matrix is diagonalized by an extended $6\times6$ unitary matrix $\mathcal{Z}$
\begin{equation*}
(\tilde{\chi}^{0}_G)^TM_{\tilde{\chi}^0}\tilde{\chi}^{0}_G=\big((\tilde{\chi}^{0}_G)^T\mathcal{Z}^T\big)\big(\mathcal{Z}^\ast M_{\tilde{\chi}^0}\mathcal{Z}^\dag\big)\big(\mathcal{Z}\tilde{\chi}^{0}_G\big)
=(\tilde{\chi}^{0}_M)^T\mbox{diag}(M_{\tilde{\chi}^0})\tilde{\chi}^{0}_M ,
\end{equation*}
\begin{equation}
(\tilde{\chi}^{0}_M)_i=\mathcal{Z}_{ij} (\tilde{\chi}^{0}_G)_i
\end{equation}
\par
Using the mixing matrices above, one may now write down the couplings of the physical charginos and neutralinos with the gauge bosons and Higgs fields.


\subsection{The $\Sigma$ Bidoublet} \label{sec:Sigma}
\par
Before discussing the phenomenological implications, let us briefly revisit the Lagrangian terms for the bidoublet $\Sigma$ to fix some of our notations and conventions.
\par
The kinetic term of the supermultiplet is
\begin{equation}\label{SigmaLagrangian}
\mathcal{L}_{\Sigma\thinspace kin}=\frac{1}{2}\mbox{Tr}\Big((D^\mu\Sigma)^\dag D_\mu\Sigma\Big)+\frac{1}{2}\mbox{Tr}\Big(\tilde{\Sigma}^\dag i \bar{\sigma}^\mu D_\mu\tilde{\Sigma}\Big)+\frac{1}{2} \mbox{Tr}F_\Sigma F_\Sigma^\ast \, ,
\end{equation}
where we have
\begin{equation}\label{Component}
\Sigma= \left(
         \begin{array}{cc}
           u+\Sigma^0+\Sigma^3 & \Sigma^1-i\Sigma^2 \\
           \Sigma^1+i\Sigma^2 & u+\Sigma^0-\Sigma^3
         \end{array}
       \right)
\end{equation}
after the breaking of $SU(2)_1\otimes SU(2)_2$. The $\Sigma^a$ corresponding to each generator are complex fields. 

The D-terms involving the bidoublet and Higgs are
\begin{equation}\label{CompleteD}
\mathcal{L}_{D,\Sigma-H}=-\frac{1}{8}g_1^2\mbox{Tr}\Big(\Sigma^\dag t^a\Sigma\Big)^2-\frac{g_2^2}{2}\Big( \frac{1}{2}\mbox{Tr}\Big(\Sigma^\dag t^a\Sigma\Big)+H^\dag_u t^aH_u+H^\dag_d t^aH_d\Big)^2.
\end{equation}
If one of the $\Sigma$ fields takes the vev $u$, we can write $\Sigma \rightarrow u + \Sigma$ with $\mbox{Tr}\Big(\Sigma^\dag t^a\Sigma\Big)\rightarrow u (\Sigma+\Sigma^{\ast})^a + {\cal{O}}(\Sigma^2)$. We may then define the massive, real triplet
\begin{equation}\label{sigmatriplet}
\Sigma_T^a \equiv \frac{1}{\sqrt{2}}(\Sigma+\Sigma^{\ast})^a\, .
\end{equation}
It follows that
\begin{equation}\label{SigmaMass}
\mathcal{L}_{D, \Sigma-H}\rightarrow-\frac{1}{4}(g_1^2+g_2^2)u^2\sum_{a=1}^3(\Sigma^a_T)^2 - \frac{g_2^2}{\sqrt{2}} u (H^\dag_u \Sigma_T^a t^a H_u + H^\dag_d \Sigma_T^a t^a H_d)-\frac{g_2^2}{2}\Big( H^\dag_u t^aH_u+H^\dag_d t^aH_d\Big)^2,
\end{equation}
and the scalar $\Sigma_T$ gets an additional mass contribution through the supersymmetric D-term. 
Combined with the mass terms from the F and soft terms of the scalar potential, the mass of the scalar triplet $\Sigma_T$ comes out to be
\begin{equation}
M^2_\Sigma =\frac{1}{2}(g_1^2+g_2^2)u^2+2m_\Sigma^2.
\end{equation} 
The Feynman rule for the $H^*$$H$$\Sigma_T^a$ vertex is $-\frac{1}{\sqrt{2}}ig^2_2ut^a$.


\section{Phenomenology of the $SU(2)_1\otimes SU(2)_2$ Gauge Extension}
\par
Let us now discuss some of the phenomenological implications of the SUSY $SU(2)_1\otimes SU(2)_2$ gauge extensions. Compared to earlier studies \cite{Batra:2003nj, Batra:2004vc,Medina:2009ey, Carena:2004ha,Craig:2012hc}, the new aspects that we would like to concentrate on are the modifications of the effective quartic coupling of the SM-like Higgs boson and the possible effects of the new charginos on the Higgs-diphoton partial decay width.


\subsection{The Enhanced D-Term Contribution to the Tree-Level Higgs Mass}
The original motivation of BDKT \cite{Batra:2003nj,Batra:2004vc} in considering $SU(2)_1\otimes SU(2)_2$ gauge extensions was to achieve large Higgs masses while avoiding the tree-level MSSM bound. This was accomplished by integrating out the heavy triplet $\Sigma$ with its equation of motion:
\begin{equation}
\Sigma_T^a = - \frac{1}{M_\Sigma^2} \frac{g_2^2 u}{\sqrt2} (H_u^\dag t^a H_u + H_d^\dag t^a H_d + \ldots) + \mathcal{O}\Big( \frac{p^2}{m_\Sigma^2} \Big) .
\end{equation}
Substituting this in Eqn.~(\ref{SigmaMass}), we have
\begin{align}
\mathcal{L}_{D, \Sigma-H} &\rightarrow -\frac{g^2 \Delta}{2} \Big( H^\dag_u t^aH_u+H^\dag_d t^aH_d\Big)^2,
\intertext{with}
\Delta &= \frac{ 1 + \frac{4m_\Sigma^2}{u^2}\frac{1}{g_1^2} }{ 1 + \frac{4m_\Sigma^2}{u^2} \frac{1}{g_1^2+g_2^2} } .
\end{align}
Compared to the MSSM, the effective $SU(2)_L$ D-term is enhanced by this factor $\Delta$. The tree-level CP-even Higgs mass matrix becomes
\begin{equation}\label{CPEvenHiggsMass}
M^2_{H^0}=\left(\begin{array}{cc}
\frac{1}{4}(g^2\Delta+g'^2)v^2\cos^2\beta+M_A^2\sin^2\beta & -(\frac{1}{4}(g^2\Delta+g'^2)v^2+M_A^2)\sin\beta\cos\beta\\
-(\frac{1}{4}(g^2\Delta+g'^2)v^2+M_A^2)\sin\beta\cos\beta & \frac{1}{4}(g^2\Delta+g'^2)v^2\sin^2\beta+M_A^2\cos^2\beta
\end{array}\right),
\end{equation}
where $g'$ is the SM $U(1)_Y$ gauge coupling and $M_A$ is the mass parameter of the CP-odd Higgs. The key observation of BDKT was that in the decoupling limit, with large $M_A$, the mass of the light, neutral, CP-even Higgs is not bounded at tree level by $M_{Z} |\cos2\beta|$, but rather by
\begin{equation}\label{TreeLevelBound}
m_h \leq \frac{1}2\sqrt{g^2\Delta+g'^2}v |\cos2\beta|.
\end{equation}
The tree-level mass splitting between the charged and CP-odd Higgs is also modified \cite{Medina:2009ey}:
\begin{equation}
m^2_{H^\pm}-m^2_{A}= \frac{g^2\Delta}{4} v^2 .
\end{equation}
Here, as before, the weak coupling constant is defined as $g=g_1 g_2/\sqrt{g^2_1+g^2_2}$.



\subsection{The Chargino Loop Contribution to the Higgs Diphoton Decay Rate}
\par
The Higgs diphoton decay is loop induced and may include contributions from bosons, fermions and scalars (see for instance \cite{Carena:2012xa, Djouadi:2005gi, Djouadi:2005gj})

\begin{equation}
\label{hgagaloop}
\Gamma(h\to \gamma\gamma)=\frac{ \alpha^2 m_h^3}{1024\pi^3}\left|\frac{g_{hVV}}{m_V^2} Q_V^2 A_1(\tau_V)+ \frac{2g_{hf\bar{f}}}{m_f} N_{c,f} Q_f^2  A_{1/2}(\tau_f) +  N_{c,S} Q_S^2 \frac{g_{hSS}}{m_S^2} A_0(\tau_S) \right |^2 \ ,
\end{equation}
where $\tau_i=4 m^2_i/m^2_h$ and $V$, $f$, and $S$ refer to spin-1, spin-1/2, and spin-0 fields. The corresponding $g_{hii}$, $Q_i$ and $N_{c,i}$ denote the coupling, electric charge and number of colors of each particle contributing to the amplitude. $A_1\,,A_{1/2}\,\text{and}\, A_{0}$ are the related loop-functions.
\par
For heavy particles in the loop, the Higgs diphoton partial width may also be quantified using Higgs low-energy theorems\cite{Ellis:1975ap,Shifman:1979eb}
\begin{align}\label{LowETheorem}
{\cal L}_{h\gamma\gamma} \simeq \frac{\alpha}{16\pi}\frac{h}{v} \frac{\partial}{\partial \log v} & \left[ \sum_i   b_{V,i} \log\left(\det {\cal M}_{V,i}^2\right)
+ \sum_i   b_{f,i} \log\left(\det {\cal M}_{f,i}^\dagger {\cal M}_{f,i}\right) \right. \nonumber \\
& \left. \qquad +\sum_i   b_{S,i}  \log\left(\det {\cal M}_{S,i}^2\right)
\right]
F_{\mu\nu} F^{\mu\nu} ,
\end{align}
where ${\cal M}_i$ are the mass matrices and $b_i$ are the coefficients of the QED one-loop beta function \cite{Carena:2012xa}.
\par
In the SM, it is well known that the dominant contribution to the amplitude is from the $W^\pm$ boson loops. For a 125 GeV Higgs boson, the loop factor $A^{\text{W}}_1$ in Eq.~(\ref{hgagaloop}) is about $-8.32$ and destructively interferes with the top-loop contribution, which gives a subdominant contribution $N_c Q^2_t A_{1/2}\simeq1.84$. 
In general, the $b_i$ coefficients of all matter particles are positive.  Hence if the determinant of the mass matrix of some new matter sector has a negative dependence on $v$, then these new particles will contribute additively to the $W^\pm$ loop and they will enhance the Higgs-diphoton partial width.
There are several different ways to achieve this that have been explored in the literature~\cite{Carena:2011aa,Djouadi:1998az, hgagaRef1,hgagaRef2, hgagaRef3, dnq, Carena:2012mw, Low:2009di, Low:2009nj, Carena:2012xa, Carena:2012gp, Joglekar:2012vc, Blum:2012ii}. In this work, we shall assume that all sfermion masses are of at least a few hundred GeV and therefore their contributions to Eq.~(\ref{hgagaloop})  is suppressed.

\par
The above situation applies in the MSSM to the charginos. Since the Higgs vev $v$ appears only in the off-diagonal entries of the mass matrix
\begin{equation}\label{CharginoMSSM}
M_{ij}^\pm=\left(\begin{array}{cc}
M_2 & \frac{1}{\sqrt{2}}gv\sin\beta \\
\frac{1}{\sqrt{2}}gv\cos\beta & \mu
\end{array}\right),
\end{equation}
we have $\det M^\pm_{ij}=M_2\mu-\frac{1}{4}g^2v^2\sin2\beta$. Therefore, in the low energy limit,
\begin{equation}\label{CharginoDiphoton}
\lim_{p\rightarrow0}\mathcal{M}(Xh)_{\text{\tiny{MSSM}}}\propto \frac{\partial}{\partial v}\log \det M^\pm_{ij} = -\frac{g^2v\sin2\beta}{2M_2\mu-\frac{1}{2}g^2v^2\sin2\beta}\simeq-\frac{g^2v\sin2\beta}{2M_2\mu},
\end{equation}
and the chargino contribution to the amplitude constructively interferes with the dominant $W^\pm$ loop to enhance the Higgs diphoton decay rate. Note that the contribution is proportional to $\sin 2 \beta$ and therefore has a maximum near $\tan \beta=1$. Unfortunately, the MSSM chargino alone cannot account for the observed enhancement~\cite{Blum:2012ii, Tao:2012pri}. This can be understood as a limitation imposed by the size the weak gauge coupling $g$. We can try to increase the effect by making the charginos lighter, but we are limited by the experimental lower bound on their masses of about 103.5 GeV at low $\tan\beta$ \cite{pdg,LEPSUSYwg}.
\par
In certain regions of parameter space of our model, the lightest chargino can have a large $\tilde{W}_2$ component. In this case, the above constraints can be overcome -- the lightest chargino couples to the Higgs with a factor enhanced by $g_2/g$ with respect to the MSSM.
The estimation of the amplitude in our model proceeds as in the MSSM case, with the added complication of the extended $4\times4$ chargino mass matrix, Eq.~(\ref{Chargino12}). For simplicity, we shall assume, that both $M_{\tilde{\Sigma}}$ and $M_{\tilde{W}_1}$ are large; therefore, at low energies, the lightest charginos are mostly admixtures of the Higgsino and the wino $\tilde{W}_2$, which couples strongly to the Higgs. The heavy charginos are then composed mostly of $\tilde{\Sigma}$ and $\tilde{W}_1$, and decoupling them introduces a seesaw-like correction to the effective $2\times2$ mass matrix of the lightest charginos:
\begin{eqnarray}\label{Chargino12eff}
M^{\pm, \text{eff}}_{ij} &\sim&\left(\begin{array}{cc}
M_{\tilde{W}_2} - \frac12 \frac{g_2^2 u^2}{M_{\tilde{\Sigma}}} - \frac{g_1^2 g_2^2}4 \frac{u^4}{M_{\tilde{\Sigma}}^2 M_{\tilde{W}_1}} & \frac{1}{\sqrt{2}}g_2vs_\beta \\
\frac{1}{\sqrt{2}}g_2vc_\beta & \mu \\
\end{array}\right),
\end{eqnarray}
where we have neglected higher-order corrections from decoupling $M_{\tilde{W}_1}$. We perform a detailed study of the rate of the Higgs decay into diphotons, the  electroweak constraints and the vacuum stability in the $SU(2)_1 \otimes SU(2)_2$ gauge extended model. To this end, we have used a modified version of the program \texttt{FeynHiggs}~\cite{Heinemeyer:1998yj} that incorporates the extended chargino and neutralino sectors.

\subsection{Radiative Corrections to the Higgs Mass and Vacuum Stability Constraints}
\par
The tree-level Higgs mass enhancement from the non-decoupling D-term is also accompanied by a potentially large loop correction. As motivated earlier, the diphoton enhancement calls for a light, strongly interacting chargino in our model. As fermions, these charginos contribute to the renormalization group equation (RGE) of the Higgs quartic coupling in a way similar to the top quark, i.e. they produce a large negative beta function contribution.
The chargino effects are small in the MSSM because the coupling $g$ is small, but they are potentially relevant in our model because their RGE contribution is proportional to $g_2^4$.
\par
This effect can be explained in two different ways. Fixing the low energy quartic coupling by the measured Higgs mass $M_h \sim 125$ GeV, the  bottom-up RGE running of  the quartic coupling drives it to negative values.  In this case, a new vacuum deeper than the physical one is generated  and the physical vacuum becomes unstable. This is generally the viewpoint adopted in non-supersymmetric models. To solve this problem, we need new bosons at some intermediate scale that stabilize the potential via positive contributions to the RGE and possible tree-level threshold corrections \cite{Carena:2004ha, EliasMiro:2012ay,Degrassi:2012ry, ArkaniHamed:2012kq}.
\par
 On the other hand, in our model, we can inversely fix the quartic coupling by Eq.~(\ref{TreeLevelBound}) at the scale $m_{W'}=\sqrt{\frac12 (g_1^2+g_2^2)}u$, where SUSY is broken for the new $W'$ and $\tilde{W}'$ sector. The quartic coupling will then be enhanced in its RG evolution to low energies via the effects of the charginos and the top quark and its supersymmetric partners.  Such effects may be strong enough to drive the Higgs mass to values larger than 125~GeV, and therefore a detailed analysis of these effects is required.
Working in the unbroken phase of the electroweak interactions and with gauge eigenstates ($\tilde{W}, \tilde{W}', \tilde{H}$), the (overall) new chargino and neutralino loop correction to the SM Higgs quartic coupling is
 ~\\
\begin{equation}\label{QuarticRGE}
\left. \frac{d \; \lambda_H }{d \log Q}  \right|_{\tilde{\chi}} =-\frac{1}{32\pi^2}\left[2g_2^4 + (g_2^2+g'^2)^2\right].  
\end{equation}
Here $\lambda_H^{\rm tree} =\frac{1}{8}(g^2\Delta+g'^2)\cos^2 2\beta$ is the tree-level quartic coupling, which can be taken as its boundary condition at the scale of the heavy gauge boson masses in the running to low energies.\footnote{We adopt the convention of $\mathcal{L}_{SM}\supset -\lambda_H (H^\dag H)^2$ and $v\sim246$ GeV.} The first (second) term on the RHS corresponds to the contribution of the chargino (neutralino). The chargino contribution is the dominant term determining the RG evolution.  We may neglect terms proportional to $\cos 2\beta$, like the ones provided by first and second generation squarks and sleptons,  since our focus will be on the regions near $\tan\beta \sim 1$ where the Higgs diphoton decay BR is maximized. Other particles that have potentially large beta function contributions include the non-standard charged and neutral Higgs bosons.  These contributions are not proportional to $\cos2\beta$, and are positive; however, their contribution is only $\frac{1}{(4\pi)^2}\frac{3}{8}g_2^4$, which is a quarter of the chargino contribution \cite{Ibrahim:2000qj}.

If we perform the calculation with mass eigenstates, the loop contribution of a chargino or a neutralino enters only if its external momentum is greater than its mass. We need to take into account the mixing matrices $\mathcal{U}$, $\mathcal{V}$ and $\mathcal{Z}$; however, as we are considering large values of $M_{\tilde{\Sigma}}$ and   $M_{\tilde{W}_1}$, significant mixing occurs only between the wino $\tilde{W}_2$ that couples strongly to the Higgs and the Higgsino, as assumed before. In this regime, we can consider approximate $2 \times 2$ unitary mixing matrices $U$ and $V$, diagonalizing Eq.~(\ref{Chargino12eff}).


\subsection{Electroweak Precision Constraints}

The electroweak precision constraints in $SU(2)_1 \otimes SU(2)_2$ gauge extended models have been extensively analysed in the literature, for instance, in Ref.~\cite{Chivukula:2003wj}. These models introduce new corrections to the electroweak precision observables induced by the tree-level mixing of the $W^\pm$ and $Z$ with their heavy counterparts $W'$ and $Z'$. These corrections are beyond the oblique ones, in which all new physics comes in the loop correction of $W^\pm$ and $Z^0$. Generally, these bosonic corrections may be parametrized by a small number of coefficients, which involve the new gauge couplings and the vacuum expectation value of the $\Sigma$ field. However, the model we are considering has a non-universal flavor structure. The first two generation chiral supermultiplets couple to one $SU(2)$ gauge group and the third generation and Higgs sector to the other $SU(2)$, and hence the electroweak precision constraints results obtained in Refs.~\cite{Chivukula:2003wj,Batra:2003nj} cannot be directly used.

The model under consideration has  common features with both case I and case II of Ref.~\cite{Chivukula:2003wj}. Observables associated with the first two generation of fermions may be treated in analogy with case II,  while those associated with  the third generation and the Higgs receive similar corrections as in  case I.  The corrections to the third generation observables are then parametrized by  the coefficient $c_1=(g/g_1)^4(v^2/2u^2)$, while due to their weaker coupling to the new gauge bosons, the first and second generation corrections are parametrized by the smaller coefficient $d_1=-(g/g_1)^2(g/g_2)^2(v^2/2u^2)$.

Because most electroweak precision measurement observables are associated with  the first two generations and the $W$ mass, naively we would expect the constraint to be dominated by similar bounds as in case II of Ref.~ \cite{Chivukula:2003wj}. However, the third generation observables, in particular for $R_b=\Gamma_b/\Gamma_{had}$, receive larger corrections, and then carry a strong weight in the global fit.  In addition, the gauge boson induced corrections to $R_b$ are large and negative, and then aggravate the deficiency of the SM theoretical prediction. This is the source of the major tension in the global fit to the precision electroweak observables. Let us mention in passing that one could in principle consider  corrections  to $R_b$ by loops including stops and  charginos.  Light, right-handed stops with masses of the order of the top quark mass are currently allowed by experimental constraints \cite{Aad:2012uu} and can lead to corrections to $R_b$ that are similar in magnitude, but of opposite sign,  to the ones induced by the heavy gauge bosons \cite{Wells:1994cu}. In the following we shall not consider this possibility and use conservatively the stronger constraints obtained in the presence of heavy stops.
\par
In addition to the corrections discussed above, there are oblique corrections that may be parametrized by the Peskin-Takeuchi $S$, $T$ and $U$ parameters.
The bidoublet $\Sigma$ gives an additional contribution to the electroweak precision observables, which may be quantified as \cite{Batra:2003nj,Batra:2004vc}
 \begin{equation}\label{DeltaT}
 \Delta T\sim\frac{4\pi}{s_W^2c_W^2}\frac{g_2^4}{g^4}\frac{M_W^2u^2}{M_\Sigma^4} \, .
 \end{equation}
Given that we shall consider large values of $M_\Sigma$, these corrections are small in our case.

The oblique corrections induced by the new charginos and neutralinos are similar to ones induced the MSSM neutralinos and charginos, with appropriate modifications to the mixing matrices and $W^\pm/Z^0$ neutralino/chargino vertices. We have therefore also taken into account the fermionic oblique corrections to the  $S,T,U$ parameters
and performed the analysis of the precision electroweak observables as outlined in \cite{Burgess:1993vc}. In order to do this, we  calculated the corrections to various electroweak precision measurements in terms of $c_1$ and $d_1$, which are summarized in Appendix \ref{EWPappendix}, and we performed a global fit to these observables~\cite{Han:2004az} using the most recent PDG data~\cite{pdg}.




\section{Results and Discussion}

\subsection{Higgs Diphoton Decay Rate}

The relevant parameters of our model are the three mass terms in the chargino/neutralino mass matrices of $M_{\tilde{W}_1}$, $M_{\tilde{W}_2}$, and $\mu$, one soft mass term in the neutralino mass matrix, $M_1$, the ratio $g_2/g_1$, the bidoublet vev $u$, $\tan\beta$, the soft mass $m_\Sigma^\prime$, and the fermionic triplet soft mass $M_{\tilde{\Sigma}}$. The top-stop sector also significantly affects the Higgs mass, and therefore the average stop mass $M_{SUSY}$ and the mixing term $\tilde{A}_t=A_t-\mu\cot\beta$ should be specified in order to compute $m_h$.

\begin{figure}
\centering
\includegraphics[height=3.4in]{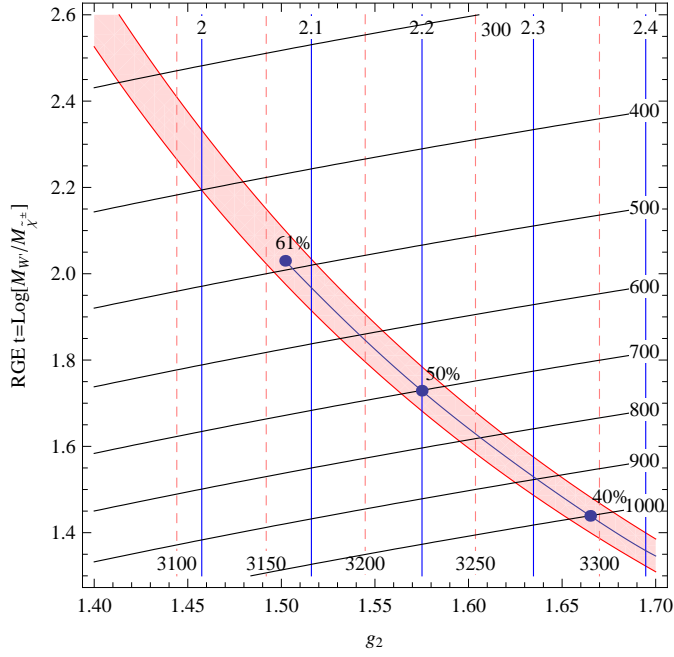}
\caption{Contours of the ratio $g_2/g_1$ (blue vertical lines), the bidoublet vev $u$ (pink dashed vertical lines) saturating the constraint $\frac{1}{2}(\frac{g}{g_1})^4(\frac{v}{u})^2 \lesssim 2\times10^{-3}$ (see Fig. \ref{EWPOfit}), and the effective chargino mass $M_{\tilde{\chi}^\pm}$ (black curves) in the plane of the $SU(2)_2$ coupling $g_2$ and the chargino/neutralino RGE running e-folding $t=\log M_{W'}/M_{\tilde{\chi}^\pm}$. Points in the red band generate a Higgs mass between 124 and 127 GeV. As explained in the text, the lowest  effective chargino mass consistent with the experimental and theoretical constraints is 490~GeV. Maximal values of the enhancement of the Higgs diphoton decay BR are denoted for some corresponding points in the red band -- the 61\% enhancement corresponds to $M_{\tilde{\chi}^\pm} = 490$ GeV. We assume soft supersymmetry breaking masses of the left- and right-handed stops of about 550~GeV and a mixing mass term $\tilde{A}_t = 500$~GeV.}
\label{RGErunning}
\end{figure}

\begin{figure}
\centering
\includegraphics[height=3.2in]{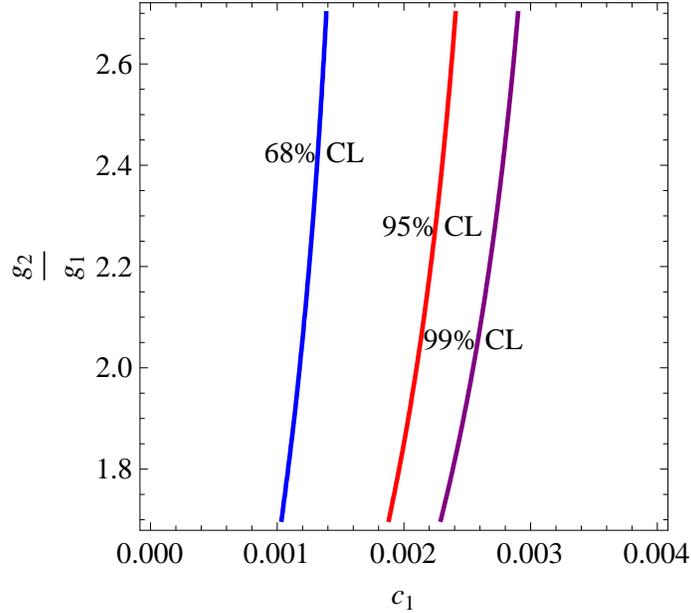}
\caption{Confidence level bounds at 68\%, 95\%, and 99\% on $c_1=\frac{1}{2}(\frac{g}{g_1})^4(\frac{v}{u})^2$ and $g_2/g_1$ based on the global fit of the electroweak precision observables. Here we have fixed $S=0.11$, $T=0.075$, as explained in the text.}
\label{EWPOfit}
\end{figure}

\par
At first glance, we see that the desired chargino contribution to the Higgs diphoton decay rate is proportional to $g_2^2$, but it is constrained by the value of the Higgs mass, determined from the Higgs quartic coupling RGE, which is proportional to $g_2^4$. In order to maximize the Higgs diphoton decay rate, we need to maximize $g_2/g_1$, choose $\sin2\beta\sim1$, and minimize $\mu M_2$, with all the relevant constraints in mind. Our benchmark scenario is $\tan\beta \sim 1$, which simultaneously maximizes the diphoton enhancement in Eq.~(\ref{CharginoDiphoton}) and minimizes the tree-level Higgs mass in Eq.~(\ref{TreeLevelBound}) to allow the maximal room for the RGE running from the chargino/neutralino sector. The Higgs mass is almost completely generated by radiative effects, so the parameter $m_\Sigma$ does not affect the tree-level Higgs mass; the only constraint pushing $m_\Sigma$ to large values comes from suppressing the contribution of the triplet to the $T$ parameter in Eq.~(\ref{DeltaT}).

Since the Higgs does not couple to $SU(2)_1$, the effect of $M_{\tilde{W}_1}$ on the diphoton enhancement and quartic coupling running is small so, as stated before, we can choose a large $M_{\tilde{W}_1}$ value to decouple $\tilde{W}_1$. This choice maximizes the mixing between the $\tilde{W}_2$ wino and the Higgsino, and hence the diphoton decay BR enhancement. In the same way, we can decouple $\tilde{\Sigma}$. We will take large values for $M_{\tilde{W}_1} = 5$ TeV and $M_{\tilde{\Sigma}}= 10$ TeV.

We can calculate the quartic coupling RGE in mass eigenstates, as discussed in Sec.~3.3: given that the heaviest and second heaviest charginos are decoupled, one may compute an effective chargino mass, $M_{\tilde{\chi}}^\pm$, parameterizing the full chargino contribution to the Higgs quartic coupling beta function, Eq.~(\ref{QuarticRGE}),
\begin{equation}
b_{\tilde{\chi}^+} \log\frac{M_{\tilde{\chi}^\pm_3}}{M_{\tilde{\chi}^\pm}}=b_1\log\frac{M_{\tilde{\chi}^\pm_2}}{M_{\tilde{\chi}^\pm_1}}+b_2\log\frac{M_{\tilde{\chi}^\pm_3}}{M_{\tilde{\chi}^\pm_2}}
\end{equation}
where $b_i$ is the beta function coefficient after mass eigenstate $i$ enters the loop, and $b_{\tilde{\chi}^+}$ is given by the right-hand side of Eq.~(\ref{QuarticRGE}). The Higgs mass at one-loop level is then dominated by the chargino-neutralino and top-stop sectors
\begin{equation}\label{HiggsMass}
M_h^2\simeq\frac{v^2}{16\pi^2} \big( 2g_2^4+(g_2^2+g'^2)^2 \big)\log\frac{M_{W'}}{M_{\tilde{\chi}^\pm}}+\frac{3v^2}{4\pi^2}y_t^4\left(\log\frac{M_{SUSY}}{M_t}+\frac{\tilde{A}_t^2}{2M_{SUSY}^2}\left(1-\frac{\tilde{A}_t^2}{12M_{SUSY}^2}\right)\right),
\end{equation}
where $y_t=\sqrt{2}m_t(M_t)/v$, where $y_t$ is the SM top Yukawa coupling at the top mass scale, with $m_t(M_t)$ the running top mass at the same scale,  $M_{SUSY}=\frac{1}{2}(M_{\tilde{t}_1}+M_{\tilde{t}_2})$ is the SUSY breaking scale for stop sector as in the usual MSSM, and $\tilde{A}_t=A_t-\mu\cot\beta$. We adopt a small top-stop sector contribution of a few thousand GeV$^2$ in order to maximize the RGE running of the chargino-neutralino sector. In our benchmark scenario, we choose $550$ GeV for the soft breaking parameters of the left-handed and right-handed stops and $\tilde{A}_t\simeq 500$~GeV for the stop mixing parameter. Including two-loop QCD and top Yukawa induced corrections, lead to a contribution to $M_h^2$ of about $5000$ GeV$^2$, compared to $(125\,\text{GeV})^2=15600$ GeV$^2$ \cite{Carena:1995wu}. The Higgs mass then gives a constraint between $g_2$ and the light chargino masses: if we increase $g_2$ in Eq.~(\ref{QuarticRGE}), the allowed RGE runnning space quantified by $\log M_{W'}/M_{\tilde{\chi}^\pm}$ is reduced, which will logarithmically raise the chargino/neutralino mass and eventually suppress the diphoton decay BR. In Fig.~\ref{RGErunning}, we have plotted with a red band the region of parameters consistent with the Higgs mass measurement.

The key constraint comes from the electroweak precision measurement. As discussed before, the bosonic sector contributions  are parameterized by the coefficients $c_1$ and $d_1$, which depend only on $u$ and $g_2/g_1$ in our parameter set. On the other hand, for  low values of $\tan\beta\sim 1$ and keeping the lightest chargino mass close to the LEP bound (as required to maximize the Higgs diphoton decay rate), the fermionic contributions to the $S$ and $T$ parameters do not present large variations. In Fig.~\ref{RGErunning}, $S \sim T \sim 0.1$ for different values of $g_2/g_1$ and $u$ lying in the red band that generate proper Higgs masses.
The region of parameters selected by the fit to precision measurement data at 95\% confidence level is shown in Fig.~\ref{EWPOfit}.\footnote{Our best fit with all five parameters has a $\chi^2/$dof of 14.9/20. If we set $T=0.075$, $S=0.11$, and $U=0$, we have a best fit of 18.8/23. In comparison, the SM is 17.2/25.} For a given value of $g_2/g_1$, from Fig. \ref{EWPOfit}, we can find the value of $u$, and hence the $W'$ mass, that saturates the bound on $c_1$. We are left with only two undetermined relevant parameters -- $M_{\tilde{W}_2}$ and $\mu$  -- in the chargino sector.
%

\begin{figure}[ht]
\begin{center}
\includegraphics[height=3.2in]{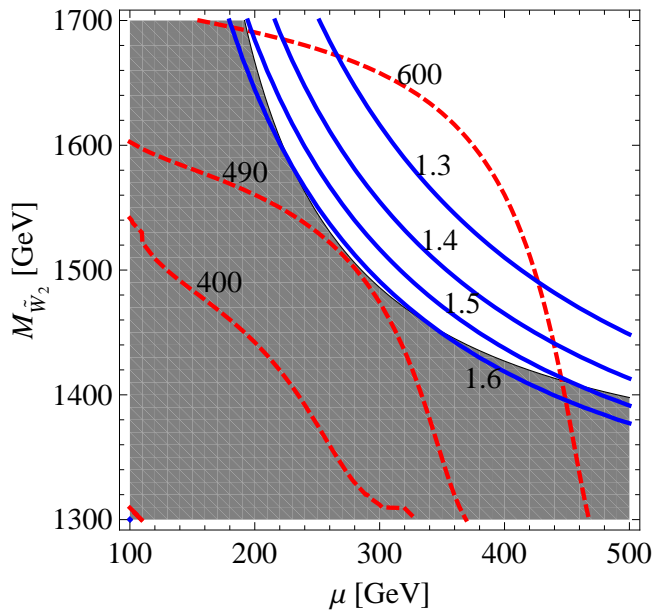}
\end{center}
\vspace{-1em}
\caption{Chargino mass and diphoton decay BR contours, for the slice with $g_2/g_1=2.08$, $u=3160$ GeV, $M_{\tilde{W}_1}=5$ TeV and $M_{\tilde{\Sigma}}=10$ TeV. The grey region is excluded by the LEP bound on the lightest chargino mass $M_{\tilde{\chi}^\pm_1}>103.5$ GeV. The diphoton decay BR enhancement contours are blue curves. The effective RGE starting scale $M_{\tilde{\chi}^\pm}$ contours are red dashed curves. At the tangent point of the 103.5 GeV lightest chargino mass bound and the $M_{\tilde{\chi}^\pm} = 490$ GeV curve, a Higgs diphoton decay BR of $3.75\times10^{-3}$, or an enhancement of about $61\%$, can be achieved.}
\label{Start353}
\end{figure}
\begin{figure}[ht]
\begin{center}
\includegraphics[height=3.2in]{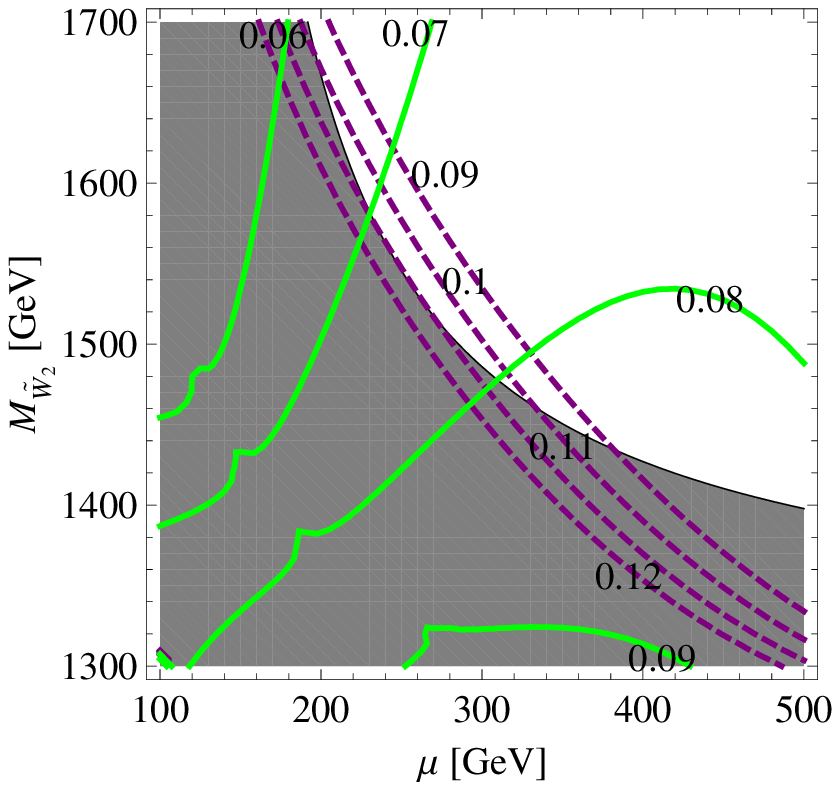}
\end{center}
\vspace{-1em}
\caption{Contours of electroweak precision observables $S$ and $T$ parameters for the same choice of model parameters as in Fig. \ref{Start353}. The chargino and neutralino contributions to the Peskin-Takeuchi $T$ and $S$ parameters are the green and purple dotted curves. We can see that at the corresponding point for maximal Higgs diphoton decay BR enhancement, $T=0.075$, $S=0.11$.}
\label{Start354}
\end{figure}

The scan results are shown in Fig.~\ref{Start353}, where we have chosen $u = 3160$~GeV and $g_2/g_1 = 2.08$. The $M_{\tilde{\chi}^\pm}$ scales are shown as the red dashed curves in Fig.~\ref{Start353}. The LEP bound on the lightest chargino of 103.5 GeV is given by the shaded gray area, while the blue lines show the enhancement of the BR of the SM-like Higgs decay into diphotons over the SM value. From Fig.~\ref{RGErunning}, we see that the effective chargino mass leading to the proper Higgs mass value for these parameters is about 490~GeV. The negative corrections to $M_{\tilde{W}_2}$ in the $(1,1)$ element of the effective chargino matrix of Eq.~(\ref{Chargino12eff}) are about 1.2 TeV. This figure therefore shows the case in which this effective chargino mass is obtained for parameters leading to maximal mixing between the wino $\tilde{W}_2$ and the Higgsino and a lightest chargino mass close to the LEP bound.\footnote{The 490 GeV value of the effective chargino mass is close to the next-to-lightest chargino mass. The lightest chargino contributes to the quartic RGE via a box diagram with four propagators fixed to be $\tilde{\chi}^\pm_1$. At energies above the next-to-lightest chargino mass, instead, both charginos can contribute via box diagrams with either $\tilde{\chi}^\pm_1$ or $\tilde{\chi}^\pm_2$ in the loop, so the effective beta function coefficient is much larger than at lower energies.} Since the contributions to the Higgs diphoton decay rate tend to be maximized for maximal mixing and for  the smallest value of the lightest chargino mass, this case therefore corresponds to the maximal value of the diphoton decay BR that may be obtained in this model. The maximal value of the Higgs diphoton decay BR is about 61\% as shown by the blue line tangent to the boundary of the grey area in the example presented in this figure. From Fig. \ref{RGErunning}, we see that larger values of $g_2$ will result in an increased effective chargino scale. Also, in spite of larger values of $g_2$, the effective coupling of the lightest chargino to the Higgs is reduced due to smaller mixing between the Higgsino and wino. 
For example, for $g_2/g_1 \simeq (2.20, 2.35)$, $M_{\tilde{\chi}^\pm} \simeq (700, 1000)$~GeV is obtained to generate the proper Higgs mass, which leads to a maximum Higgs diphoton decay BR enhancement of about $(50\%, 40\%)$. These maximal values are shown by blue dots in Fig.~\ref{RGErunning}.


In Fig.~\ref{Start354} we show the chargino and neutralino contributions to the
electroweak precision measurement $S$ and $T$ parameters, where we have also included the small contributions associated with the stop sector. The total contributions to $S$ and $T$ are small and positive, and remain consistent with the allowed values of these parameters obtained from a fit to the electroweak precision data~\cite{pdg}. We can see that at the corresponding point for maximal Higgs diphoton decay BR enhancement, $T=0.075$, $S=0.11$. 

As previously emphasized, in our benchmark scenario, we have employed light top squarks with masses of about 550 GeV in order to minimize the value of the effective chargino mass $M_{\tilde{\chi}^\pm}$. Such light stops enhance the gluon fusion rate by about 10\% compared with the SM, and provide an additional enhancement to all Higgs production rates in the gluon fusion channel. Larger stop masses will reduce this rate enhancement, but due to their impact on the Higgs mass, they will reduce the possibility of having light charginos with strong coupling to the Higgs, as is assumed in this work.  Therefore, a prediction of this model would be a slight enhancement of the gluon fusion Higgs production channels compared to the SM ones. No such enhancement should be observable in the weak boson fusion channels, apart from the obvious case of the diphoton decay rate, that was analysed in detail in this work.

In the MSSM, for $\tan\beta\lesssim1.2$, the top Yukawa coupling is large and in the RGE running blows up below the GUT scale. However, in our model, there are additional strong $SU(2)_2$ gauge coupling effects that induce a large negative contribution to the top Yukawa RGE. Using the modified RGE evolution of the gauge and Yukawa couplings, and taking into account the breaking of $SU(2)_1\otimes SU(2)_2 \rightarrow SU(2)_L$ in a consistent way \cite{Batra:2004vc},  we have checked that the top Yukawa coupling remains perturbative up to the scales of the order of the Planck scale.

Relaxing the exact condition $\tan\beta=1$ increases the tree-level contribution to the Higgs mass, Eq.~(\ref{TreeLevelBound}), and therefore reduces the possible chargino contributions to the running of the Higgs quartic coupling RGE. 
The tree-level contribution depends on the value of $\Delta$, which in turn depends on $m_\Sigma$.  Choosing values of the scalar (and fermion) triplet mass to be of the order of the heavy gauge boson masses,  $M_{\tilde{\Sigma}} \simeq m_{\Sigma} \simeq m_{W'}$, one obtains a value of $\Delta \simeq 1 + 2(g_2/g_1)^2/3$, while choosing as before large values for these masses leads to $\Delta \simeq 1 + (g_2/g_1)^2$. Therefore, the value of $\Delta$  is correlated with $g_2/g_1$.  Varying these masses in that range and taking all other parameters to 
saturate the bounds on the precision electroweak measurements and on the chargino mass of about 103.5~GeV, for $\tan\beta = 1.5$ we can get a maximal Higgs diphoton decay branching ratio enhancement of about 42\%,  while for $\tan\beta = 2.0$ the maximal enhancement reduces to about  34\%.


\subsection{Chargino and Neutralino Searches at the LHC}

\begin{figure}[ht]
\centering
\includegraphics[height=2.8in]{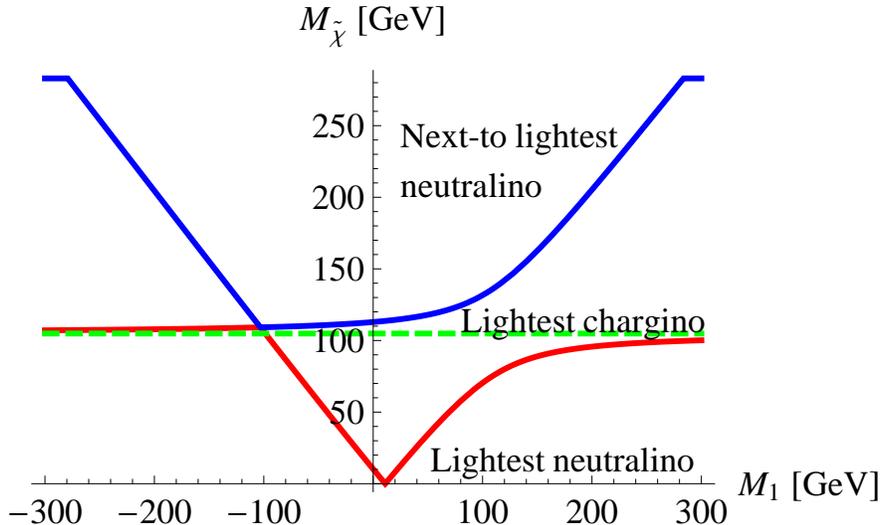}
\caption{Masses of the lightest neutralino (red curve) and the next-to-lightest neutralino (blue line) as a function of $M_1$. Also plotted is the mass of the lightest chargino (green dashed line).}
\label{MassSplitting}
\end{figure}

In the neutralino sector, in addition to the parameters entering in the chargino sector, there is a new parameter, representing the bino soft mass $M_1$ which controls the mixing between the bino, the light Higgsino, and the $\tilde{W}_2$ wino states. If the bino mass $M_1$ is sent to very large values, the lightest chargino becomes degenerate in mass with the lightest neutralino, and the chargino decay width is suppressed. This suppression may be sufficiently strong to lead to long-lived charginos. These can generate charged tracks that eventually become missing energy signatures,
which are strongly constrained. This situation may be avoided if the mass splitting is larger than a few hundred MeV \cite{Gunion:2001fu}
and from Fig.~\ref{MassSplitting}, we see that this is easily achieved if $M_1$ is smaller than a few hundred GeV.

We also plot the region of negative $M_1$ in Fig.~\ref{MassSplitting}. For small values of $|M_1|$ the lightest neutralino mass is mostly bino. The structure around $M_1 = 0$ is due to the sign change of the lightest neutralino mass.  For sufficiently negative $M_1 \lesssim -100$ GeV, the lightest chargino is even lighter than the neutralino, and they are approximately degenerate in mass. This situation is viable if the lightest chargino decays, for instance in gauge mediation with a gravitino LSP. With this spectrum, in the process $pp \rightarrow W^+ \rightarrow \tilde{\chi}^+_1\tilde{\chi}^0_1$, the neutralino can decay to a chargino with a very soft $W^\pm$ which will escape detection. Although the phase space of this process is suppressed, the competing gravitino decay mode of the neutralino is also suppressed, so the BR of its decay into charginos tends to be sizable. The two light charginos obtained in this process, as well as those directly produced in pairs, give two $W^\pm$ bosons and a small amount of MET. Therefore, charginos contribute to the total inclusive $W^+W^-$ cross section, and as suggested in Ref.~\cite{Curtin:2012nn}, may lead to a possible improvement of the agreement between the SM theoretical prediction and the experimental value of the $W^+W^-$ production rate~\cite{ATLASWW,CMSWW1, CMSWW2}.

Another constraint for the chargino/neutralino mass spectrum comes from trilepton searches~\cite{Chatrchyan:2012mea, :2012ewa, :2012gg, :2012ku, :2012ATLAStrilep}, which include the process $pp \rightarrow W^\pm \rightarrow \tilde{\chi}^\pm_1 \tilde{\chi}^0_2 \rightarrow (W^\pm \tilde{\chi}^0_1) (Z \tilde{\chi}^0_1)$, where $W^\pm$ and $Z^0$ decay leptonically. Small values $|M_1| \lesssim 50$ GeV are in tension with ATLAS chargino search results \cite{:2012ATLAStrilep}.
Note that we chose $\mu = 215$~GeV and values of $M_{\tilde{W}_2}$ such that the lightest chargino mass is close to the LEP bound. For $\tan\beta\sim1$, there is a neutralino mass eigenstate that is almost purely Higgsino, with mass very close to $\mu = 215$ GeV according to our choice of parameters. This eigenstate will be the second-lightest one if $M_1$ becomes large. 


\section{Conclusion}

In this paper, we have revisited the   $SU(2)_1\otimes SU(2)_2$ gauge extension of the MSSM in the context of the recent discovery at the LHC of a 125 GeV Higgs and measurement of an enhanced Higgs diphoton decay rate.  The $SU(2)_1 \otimes SU(2)_2$ gauge group is broken to the SM $SU(2)_L$ gauge group at TeV scale energies, where at higher energies the Higgs and the third generation quarks couple only to the strongly coupled $SU(2)_2$ gauge interactions.  The model is reviewed in some detail, making emphasis on the possible impact of light charginos and neutralinos on the predicted value of the SM-like Higgs mass. The RGE of the effective quartic coupling of the SM-like Higgs may receive important contributions from the $SU(2)_2$ charginos and neutralinos, and the same charginos can lead to an important modification of the rate of the decay of the Higgs into diphotons.

We showed that in order to maximize the effect of the light charginos on the diphoton decay rate of the Higgs, the value of $\tan\beta \sim 1$, which implies that the Higgs mass becomes small at tree-level and is mostly obtained by radiative corrections.  In the MSSM, such large radiative corrections can only be generated by very heavy stops with masses of the order of tens of TeV. In this model, however, the chargino effects may be strong enough to lift the Higgs mass to the required values even for relatively small values of the third generation squarks. Indeed, the Higgs mass puts an upper bound on the possible values of the strong $SU(2)_2$ gauge coupling and on the effective chargino mass, and therefore on its effects on the Higgs diphoton decay width. We made a detailed analysis of the precision electroweak constraints on the parameters of this model, which lead to a bound on the heavy gauge boson masses that depend on the precise value of the $SU(2)_1$ and $SU(2)_2$ couplings. We also took into account the current experimental bound on the sparticle masses. After all constraints were taken into account, we showed that an enhancement of the Higgs diphoton decay rate of about 61\% of the SM value may be obtained in this model, with perturbativity maintained up to scales of the order of the Planck scale.

In this work, we have made specific choices for the masses of the non-standard Higgs bosons as well as for the third generation squarks. We have also not considered the possible chargino two-loop effects on the RGE of the Higgs quartic coupling.  Smaller values of the CP-odd Higgs mass $m_A$ lead to a slower RGE of the Higgs quartic coupling, while due to logarithmic enhancements two-loop chargino effects may vary the Higgs mass by amounts of the same order as the one-loop heavy Higgs ones.  We reserve for future work a detailed analysis of the possible two-loop corrections as well as variations of our results with changes of the squark and Higgs mass parameters.


\section{Acknowledgments}\label{Acknowledgments}

We are grateful to Tim Tait for discussions on the EW precision measurement constraints. R.~H. wishes to thank Pedro Schwaller for useful discussions. A.~T. would like to thank Kfir Blum, Andrey Katz and Scott Thomas for discussions. G.~L. was supported by an NSERC PGS Fellowship. Work at ANL is supported in part by the U.S. Department of Energy under Contract No. DE-AC02-06CH11357. A.~T. acknowledges support from DOE grant DE-FG02-96ER40959 and the Sidney Bloomenthal fellowship during early stages of this work.


\appendix

\pagebreak

\section{Table of Electroweak Precision Constraints} \label{EWPappendix}

\begin{table}[h]
\begin{center}
\begin{tabular}{cl}
Observable & $SU(2)_1 \times SU(2)_2$ value \\

\hline

$\Gamma_Z$ & $(\Gamma_Z)_{SM}(1-0.557c_1-0.338d_1-0.00385S+0.0106T)$ \\

$R_e$ & $(R_e)_{SM}(1-0.503c_1+0.585d_1-0.00301S+0.00214T)$ \\

$R_\mu$ & $(R_\mu)_{SM}(1-0.503c_1+0.585d_1-0.00301S+0.00214T)$ \\

$R_\tau$ & $(R_\tau)_{SM}(1-1.554c_1+1.637d_1-0.00301S+0.00214T)$ \\

$\sigma_h$ & $(\sigma_h)_{SM}(1-0.611c_1+0.629d_1+0.000665-0.000700T)$ \\

$R_b$ & $(R_b)_{SM}(1-1.786c_1+1.767d_1+0.000985S-0.000472T)$ \\

$R_c$ & $(R_c)_{SM}(1+0.503c_1-0.468d_1-0.00129S+0.000913T)$ \\

$A^e_{FB}$ & $(A^e_{FB})_{SM}+0.173d_1-0.00632S+0.00449T$ \\

 $A^\mu_{FB}$ & $(A^\mu_{FB})_{SM}+0.173d_1-0.00632S+0.00449T$ \\

$A^\tau_{FB}$ & $(A^\tau_{FB})_{SM}-0.201c_1+0.374d_1-0.00632S+0.00449T$ \\

$A_\tau(P_\tau)$ & $(A_\tau(P_\tau))_{SM}-1.817c_1+2.597d_1-0.0286S+0.0203T$ \\

$A_e(P_\tau)$ & $(A_e(P_\tau))_{SM}+0.780d_1-0.0286S+0.0203T$ \\

$A^b_{FB}$ & $(A^b_{FB})_{SM}-0.015c_1+0.530d_1-0.0189S+0.0134T$ \\

$A^c_{FB}$ & $(A^c_{FB})_{SM}+0.399d_1-0.0146S+0.0104T$ \\

$M_W^2$ & $(M_W^2)_{SM}(1 + 0.430d_1 - 0.00727S + 0.112T + 0.00846U)$ \\

$g_L^2(\nu N \rightarrow \nu X)$ & $(g_L^2(\nu N \rightarrow \nu X))_{SM} + 0.246d_1 - 0.00269S + 0.00663T$ \\

$g_R^2(\nu N \rightarrow \nu X)$ & $(g_R^2(\nu N \rightarrow \nu X))_{SM} - 0.085d_1 + 0.000937S - 0.000192T$ \\

$g_{eV}(\nu e \rightarrow \nu e)$ & $(g_{eV}^2(\nu e \rightarrow \nu e))_{SM} - 0.661d_1 + 0.00727S - 0.00546T$ \\

$g_{eA}^2(\nu e \rightarrow \nu e)$ & $(g_{eA}^2(\nu e \rightarrow \nu e))_{SM} - 0.00391T$ \\

$Q_W(\text{Cs})$ & $(Q_W(\text{Cs}))_{SM} + 72.7 d_1 - 0.796S - 0.0113T$
\end{tabular}
\end{center}
\label{EWPtable}
\caption{Table of changes to electroweak precision observables in terms of the Peskin-Takeuchi parameters $S, T,U$ and the quantities $c_1=(g/g_1)^4(v^2/2u^2)$ and $d_1=-(g/g_1)^2(g/g_2)^2(v^2/2u^2)$. This is modelled on Table 3 of \cite{Chivukula:2003wj}, and the analysis follows \cite{Burgess:1993vc}.}
\end{table}


\end{document}